\documentclass[aps,prd,showpacs,groupedaddress,floatfix,nofootinbib]{revtex4}

\usepackage{amsmath}
\usepackage{graphicx}
\usepackage{color}

\newcommand{\beq}{\begin{eqnarray}}
\newcommand{\eeq}{\end{eqnarray}}

\begin{document}

\title{{Gravitational lensing from compact bodies: analytical results for strong and weak deflection limits}}

\author{Paolo Amore}
\email{paolo@ucol.mx}
\affiliation{Facultad de Ciencias, Universidad de Colima,\\
Bernal D\'{i}az del Castillo 340, Colima, Colima,
Mexico}
\author{Mayra Cervantes}
\email{mayradcv@ucol.mx}
\affiliation{Facultad de Ciencias, Universidad de Colima,\\
Bernal D\'{i}az del Castillo 340, Colima, Colima,
Mexico}

\author{Arturo De Pace}\email{depace@to.infn.it}
\affiliation{Istituto Nazionale di Fisica Nucleare, Sezione di Torino, \\
via P.Giuria 1, I-10125 Torino, Italy}

\author{Francisco M. Fern\'andez}\email{fernande@quimica.unlp.edu.ar}
\affiliation{INIFTA (Conicet,UNLP), Divisi\'on Qu\'imica Te\'orica, \\ Diag. 113 y 64 S/N,
Sucursal 4, Casilla de Correo 16, \\ 1900 La Plata, Argentina}

\begin{abstract}
We develop a non--perturbative method that yields analytical expressions for the 
deflection angle of light in a general static and spherically symmetric  metric. 
The method works by introducing in the problem an artificial parameter, called $\delta$, 
and by performing an expansion in this parameter to a given order. The results obtained
are analytical and non-perturbative because they do not correspond to a polynomial
expression in the physical parameters. Already to first order in $\delta$ the
analytical formulas obtained using our method  provide at the same time
accurate approximations both at large distances (weak deflection limit) and 
at distances close to the photon sphere (strong deflection limit). 
We have applied our technique to different metrics and verified that the error
is at most $0.5 \%$ for {\sl all} regimes. We have also proposed an alternative 
approach which provides simpler formulas, although with larger errors.
\end{abstract}


\maketitle

\section{Introduction}
\label{intro}

The theory of General Relativity (GR) predicts that  massive bodies deform the
space--time around them; even massless particles, like photons, will therefore
feel the gravitational force and their trajectories will necessarily depart 
from a straight line. Such effect is particularly strong in the proximity
of black holes, which are very massive and compact objects: in fact at a certain
distance from the black hole (the photon sphere) the deflection angle of the
photon becomes infinite. This regime is known as strong deflection limit (SDL), whereas
the regime corresponding to deflection at large distances is known as weak deflection
limit (WDL).

The WDL has been studied in a series of articles for different metrics:
for example the lensing from a Kerr metric has been considered in  
\cite{Bray:1985ew, Sereno:2003kx,Sereno:2006ss} whereas the lensing from a Reissner-Nordstrom
metric has been considered in \cite{Sereno:2003nd, Keet05}. 

In recent times there has also been great interest in studying the effects of strong
deflection limit, i.e. lensing due to light passing very close to a compact and
massive body. For example SDL in a Schwarzschild black hole
has been considered by Frittelli, Kling and Newman~\cite{FKN00} and by Virbhadra
and Ellis~\cite{VE00}; Virbhadra and collaborators have also treated the
SDL by naked singularities~\cite{VE02} and in the presence of a scalar
field~\cite{VNC98}; Eiroa, Romero and Torres~\cite{Eir02} have described Reissner-Nordstr\"om
black hole lensing, while Bhadra has considered the gravitational lensing due to the
GMGHS charged black hole~\cite{Bha03}; Bozza has studied the gravitational lensing by a
spinning black hole~\cite{Boz03}; Whisker~\cite{Whi05} and Eiroa~\cite{Eir05} have
considered SDL by a braneworld black hole; still Eiroa~\cite{Eir06}
has recently considered the gravitational lensing by an Einstein-Born-Infeld
black hole; Sarkar and Bhadra have studied the SDL in the
Brans-Dicke theory~\cite{SB06}; Konoplya has studied the corrections to the deflection
angle and time delay for a black hole immersed in a uniform magnetic field~\cite{Kono06};
Gyulchev and  Yazadjiev have studied the SDL for a Kerr-Sen dilaton 
axion black hole~\cite{Gyulchev:2006zg};
finally Perlick~\cite{Perl04} has obtained an exact gravitational lens equation
in a spherically symmetric and static spacetime and used it to study lensing by a
Barriola-Vilenkin monopole and by an Ellis wormhole. 
Notice that Bozza and Sereno~\cite{BS06} have also investigated the SDL of gravitational 
lensing by a Schwarzschild  black hole embedded in an external gravitational field. 

We can distinguish between two different approaches that have been developed to obtain
analytical expressions for the deflection angle in the strong regime: one which looks
for improvements of the weak lensing expressions, whose range of validity is therefore
extendend to distances closer to the photon sphere, without however taking into account
the divergence of the deflection angle on the photon sphere, and one which treats exactly 
the singularity of the photon sphere and whose precision rapidly drops at larger 
distances~\footnote{For a detailed discussions on the photon surface the reader can refer to
\cite{CVE}.}.
In the first category falls the work of  Mutka and M\"ah\"onen~\cite{Mutka} and of
Belorobodov~\cite{Belo02} who worked out improved formulas for the deflection angle in a
Schwarzschild metric, and the  more systematic approach of  Keeton and Petters~\cite{Keet05}
who have developed a formalism for computing corrections to lensing observables in a static
and spherically symmetric metric beyond the WDL. 
In the second category falls the work of Bozza, who has introduced an analytical method
based on a careful description of the logarithmic divergence of the deflection angle which
allows one to discriminate among different types of black holes~\cite{Boz02}.
Recently, Iyer and Petters~\cite{IP06} have also developed an analytic perturbation framework 
for calculating the bending angle of light rays traversing the field of a Schwarzchild black 
hole, obtaining accurate expressions even in proximity of the photon sphere.

In a different category falls a method developed by Amore and collaborators~\cite{AA06,AAF06}.
This method enables one to convert the integral for the deflection angle in a static and 
spherically symmetric metric into a series in an artificial parameter $\delta$.
Such series has an exponential convergence rate and its terms  
can be calculated analytically. The method is non--perturbative in the sense that it provides 
non--polynomial expressions in terms of the chosen physical parameter and yields sufficiently 
accurate  results even at first order. In our previous works we have tested our formalism on 
a variety of different metrics, always  obtaining very encouraging results.

The purpose of this paper is to improve our method in order to provide an accurate
treatment close to the photon sphere, even at first order, but without the customary deterioration
of the results at larger distances.

The paper is organized as follows: in Section~\ref{sec:method} we illustrate the application
of our method by means of the Schwarzschild metric and later we show how to treat
a general case. In Section~\ref{gm} we extend our analysis to a general static and spherically 
symmetric metric.
In Section~\ref{sec:results} we
compare our approximations with exact results and other approaches available
in the literature. Finally, in the last section we briefly summarize and discuss our results
and consider further developments.

\section{Formalism}
\label{sec:method}

Let us first review the method of Amore and collaborators~\cite{AA06,AAF06}. 
We are interested in the general static and spherically symmetric metric which corresponds to
the line element (in the following we set the velocity of light $c=1$)
\beq
\label{eq_1}
ds^2 = B(r) dt^2 - A(r) dr^2 - D(r) r^2 \left(d\theta^2+ \sin^2\theta\ d\phi^2\right) 
\eeq
and which contains the Schwarzschild metric as a special case. We also assume that the flat 
spacetime is recovered at infinity, i.e. that $\lim_{r\rightarrow \infty} f(r)  = 1$, where 
$f(r) = (A(r),B(r),D(r))$~\footnote{This is a sufficient but not necessary condition since it 
warrants that $f(r)$ is analytic around $r=\infty$; for example in \cite{Am05b} we have applied 
the method to the Weyl metric which is not asymptotically flat.}.

The angle of deflection of light propagating in this metric can be expressed by means of the 
integral~\cite{VNC98}
\beq
\Delta\phi = 2 \int_{r_0}^\infty \sqrt{A(r)/D(r)} 
\sqrt{ \left[\left( \frac{r}{r_0}\right)^2 \frac{D(r)}{D(r_0)} 
\frac{B(r_0)}{B(r)}-1\right]^{-1}} \frac{dr}{r} - \pi \ ,
\eeq
where $r_0$ is the distance of closest approach of the light to the center of the gravitational attraction.  

By introducing the variable $z=r_0/r$ one can rewrite the equation for the deflection angle as
\begin{equation}
  \Delta\phi = 2 \int_0^{1} \frac{dz}{\sqrt{V(1)-V(z) }}-\pi,
\label{eq_1_5}
\end{equation}
where 
\begin{equation}
  V(z) \equiv z^2 \frac{D(r_0/z)}{A(r_0/z)} - \frac{D^2(r_0/z)
    B(r_0)}{A(r_0/z) B(r_0/z) D(r_0)} +  \frac{B(r_0)}{D(r_0)}
\end{equation}
is a sort of ``potential'' built out of the metric. Notice that the integral in Eq.~(\ref{eq_1_5}) can be solved 
analytically only in a limited number of cases, such as for the Schwarzschild metric \cite{Darw59} and for
the Reissner-Nordtr\"om metric \cite{Eir02}, where it can be expressed in terms of elliptic integrals. 
{\sl No analytical formula can can be obtained in the case of a general static and 
spherically symmetric metric}.

Since an explicit calculation of the integral is not possible, we interpolate
the actual potential $V(z)$ with a simpler potential $V_0(z)$, which should 
be chosen in such a way that the integral~(\ref{eq_1_5}) can be performed explicitly  
when $V(z)$ is replaced with $V_0(z)$. Then we write
\begin{equation}
  V_\delta(z) \equiv V_0(z) + \delta (V(z)-V_0(z)) ,
\end{equation}
where $\delta$ is a dummy parameter. In general $V_0(z)$ may depend upon one or more arbitrary
parameters; for the time being we simply choose $V_0(z) = \lambda z^2$.
We can rewrite the expression for the deflection angle as
\begin{equation}
  \Delta\phi_\delta = 2 \int_0^{1} \frac{dz}{\sqrt{V_0(1) -V_0(z)}}
    \frac{1}{\sqrt{1 + \delta \Delta(z) }}-\pi ,
\label{eq_1_7}
\end{equation}
where
\begin{equation}
  \Delta(z) \equiv \frac{V(1)-V(z)}{V_0(1)-V_0(z)}-1 .
\label{delta}
\end{equation}
Notice that Eq.~(\ref{eq_1_7}) {\sl reduces to } Eq.~(\ref{eq_1_5}) for $\delta=1$ and therefore
is not an approximation. The expansion of Eq.~(\ref{eq_1_7}) in powers of $\delta$ converges at $\delta=1$
provided that $|\Delta(z)|<1$ for $0\leq z \leq1$. As discussed in earlier papers~\cite{Am05a,Am05b,AA06} 
this condition  requires that $\lambda$ be greater than a critical value $\lambda_C$. 
In that case one obtains a parameter--dependent series that converges towards the exact result which is however 
independent of $\lambda$. The artificial dependence on $\lambda$ observed in the  partial sums 
$\Delta\phi^{(N)}$, $N=1,2,\dots$, is minimized by means of the Principle of Minimal Sensitivity 
(PMS)~\cite{Ste81}, which corresponds to imposing the condition
\begin{equation}
  \frac{\partial}{\partial \lambda} \Delta\phi^{(N)} = 0 .
\label{eq_1_8}
\end{equation}
A proof of convergence of the series and an estimate of its rate of convergence are given
elsewhere~\cite{AA06}. 

One might be tempted at this point to question our definition of the method as being non--perturbative: 
after all, the method works by performing a perturbative expansion in
$\delta$. However, one should understand that the solution of Eq.~(\ref{eq_1_8}) is in general a function 
of the parameters in the problem and when substituted back in the series it provides non--polynomial expressions 
in the physical parameters~\cite{AA06,AAF06}, whereas the dependence upon $\delta$ desappears because it is set to one at the end of the calculation.

We are now ready to generalize this method. The first step is to write the
integral as
\begin{eqnarray}
  \Delta\phi &=& 2 \int_0^{\sigma} \frac{dz}{\sqrt{V(1)-V(z) }} +2
    \int_{\sigma}^1 \frac{dz}{\sqrt{V(1)-V(z) }}-\pi \nonumber \\
&\equiv& \Delta\phi_{a} + \Delta\phi_{b} -\pi ,\label{eq:split}
\end{eqnarray}
where $\sigma \in (0,1)$ is an arbitrary point in the region of integration.
The two integrals in this equation will now be approximated following two different
strategies.
Clearly, the particular case $\sigma = 1$ corresponds to the method just
outlined~\cite{AA06,AAF06}.

\begin{figure}
\begin{center}
\includegraphics[width=9cm]{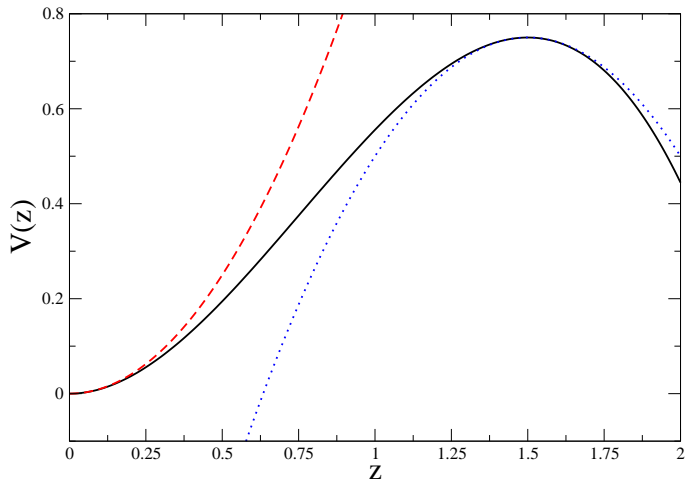}
\bigskip
\caption{ The potential $V(z)$ in Eq.~(\protect\ref{VSch}) with
  $\mu = 3/2$ (solid line). The dashed and dotted lines correspond to quadratic
  Taylor expansions around $z=0$ and $z=\mu$, respectively. (color online)}
\label{FIG1}
\end{center}
\end{figure}

For clarity,  we confine ourselves, for the moment being, to the case of the
Schwarzschild metric and later we generalize our results to arbitrary
metrics. The Schwarzschild metric is given by
\begin{equation}
  B(r) = A^{-1}(r) = \left(1-\frac{2GM}{r}\right) , \quad D(r) = 1 ,
\label{eq_2_1}
\end{equation}
and the potential $V(z)$ reads
\begin{equation}
  V(z) = z^2 - \frac{2}{3\mu} z^3 ,
\label{VSch}
\end{equation}
where $\mu \equiv r_0/3GM \ge 1$. Fig.~\ref{FIG1} shows the potential
$V(z)$ with $\mu=3/2$. The dashed and dotted lines
correspond to the quadratic Taylor polynomials around $z=0$ and $z=\mu$.
Viewed from the perspective of a classical mechanics problem, the points $z=0$
and $z=\mu$ correspond to a stable and an unstable point of equilibrium,
respectively; the nonlinear pendulum, for example, is a simple
physical system that displays this behavior. Since the integral that we want
to calculate is restricted to $z \leq 1$, the unstable point of equilibrium
will not be reached unless $\mu = 1$.
In such a case the integral will diverge and it will correspond to the photon
sphere\footnote{Notice that $\mu = 1$ yields the well--known result
$r_0 = 3GM$.}.

One expects $\sigma \approx 1$ to be a reasonable choice when $\mu \gg 1$, since
it has given accurate results earlier~\cite{AA06,AAF06}. On the other hand, as
$\mu \rightarrow 1^+$ the second integral will become increasingly more important
and one expects the optimal value of $\sigma$ to move to the center of the
integration region. We will see that this is the case later on when we discuss
a systematic way of partitioning the integral.

Our strategy is simple and consists of treating the integral in each region
differently. In the first region we follow essentially the earlier
procedure~\cite{AA06,AAF06} with the interpolating potential
$V_a(z) = \lambda z^2$, $\lambda>0$. In the second
region, on the other hand, we will interpolate the potential with the inverted
parabola $V_b(z) = V(\mu) - \rho (z-\mu)^2$, where $\rho >0$ is another arbitrary
parameter. Notice therefore that we are working with three arbitrary parameters,
$\lambda$ and $\rho$ which enter in the definition of the interpolating potentials,
and $\sigma$ which defines the border between the two regions.

In the first region we write
\begin{equation}
  \Delta\phi_a = 2 \int_0^{\sigma} \frac{dz}{\sqrt{V_a(1) -V_a(z)}}
    \frac{1}{\sqrt{1 + \delta \Delta_a(z) }} ,
\end{equation}
where
\begin{equation}
  \Delta_a(z) \equiv \frac{V(1)-V(z)}{V_a(1)-V_a(z)}-1 = -\frac{3 (\lambda -1)
    \mu  (z+1)+2 \left(z^2+z+1\right)}{3 \lambda \mu (z+1)} .
\end{equation}
After expanding to first order in $\delta$ we obtain
\begin{equation}
  \Delta\phi_a^{(1)} = \frac{2}{\sqrt{\lambda}} \int_0^{\sigma}
    \frac{dz}{\sqrt{1-z^2}} \left[ 1 - \frac{ \Delta_a(z)}{2} \right] .
    \label{eq:Da1}
\end{equation}
Straightforward integration yields
\begin{equation}
  \Delta\phi_a^{(1)} = \frac{-2 \sigma  \left(\sqrt{1-\sigma ^2}-2\right)+3 (3
    \lambda -1) \mu  (\sigma +1) \arcsin(\sigma )-4 \sqrt{1-\sigma ^2}+4}{3
    \lambda ^{3/2} \mu (\sigma +1)} .
\end{equation}
The PMS (see Eq.~(\ref{eq_1_8})) gives us the optimal value of $\lambda$
\begin{equation}
  \lambda_\text{PMS} = \frac{2 \sigma  \left(\sqrt{1-\sigma ^2}-2\right)+4
    \left(\sqrt{1-\sigma^2}-1\right) + 3 \mu  (\sigma +1) \arcsin(\sigma )}{3
    \mu  (\sigma +1) \arcsin(\sigma )}
\end{equation}
and
\begin{equation}
  \Delta\phi_a^{(1)} = \frac{2\arcsin(\sigma )}{\sqrt{
    \frac{\displaystyle 2 \left[(\sigma +2) \sqrt{1-\sigma^2}-2 (\sigma +1)
    \right]}{\displaystyle 3 \mu  (\sigma +1) \arcsin(\sigma )}+1}} .
\end{equation}
In the second region we have
\begin{equation}
  \Delta\phi_b = 2 \int_{\sigma}^1 \frac{dz}{\sqrt{V_b(1) -V_b(z)}}
    \frac{1}{\sqrt{1 + \delta \Delta_b(z) }} ,
\end{equation}
where
\begin{equation}
  \Delta_b(z) \equiv \frac{V(1)-V(z)}{V_b(1)-V_b(z)}-1 = \frac{-6 \rho \mu^2
    +3 (\rho +1) (z+1) \mu -2 \left(z^2+z+1\right)}{3 \mu \rho (2 \mu -z-1)} .
\end{equation}
After expanding to first order and integrating we obtain
\begin{equation}
  \Delta\phi_b^{(1)} = \frac{2 \left(\xi ^2+\mu -\mu \sigma +\sigma
    -1\right)+3 \mu (3 \rho -1) \xi \ln \left(\frac{\displaystyle 1-\mu }
    {\displaystyle -\mu +\sigma +\xi}\right)}{3 \mu  \rho ^{3/2} \xi } ,
\end{equation}
where we have defined
\begin{equation}
  \xi \equiv \sqrt{\sigma ^2-2 \mu  \sigma +2 \mu -1} .
\end{equation}
It is worth noticing that this simple first--order approximation exhibits the
correct logarithmic singularity at $\mu = 1$.

The PMS gives us again the optimal value of $\rho$,
\begin{equation}
  \rho_\text{PMS} = 1+\frac{2 (-3 \mu +\sigma +2) \xi }{3 \mu  (2 \mu -\sigma -1)
    \ln \left(\frac{\displaystyle 1-\mu }{\displaystyle -\mu +\sigma +\xi }
    \right)},
\end{equation}
and
\begin{equation}
  \Delta\phi_b^{(1)} = \frac{2 \sqrt{3} \ln\left(\frac{\displaystyle 1-\mu }
    {\displaystyle -\mu +\sigma+\xi}\right)}{\sqrt{\frac{\displaystyle 2
    (-3 \mu +\sigma +2)\xi }{\displaystyle \mu (2 \mu -\sigma
    -1) \ln\left(\frac{1-\mu }{-\mu +\sigma +\xi }\right)}+3}} .
\end{equation}
By adding the two expressions we obtain
\begin{eqnarray}
  \Delta\phi_\text{PMS}^{(1)} &=& \frac{2 \sqrt{3} \sqrt{\mu (\sigma +1)}
    \arcsin^{3/2}(\sigma )}{\sqrt{2 \sigma \left(\sqrt{1-\sigma^2}
    -2\right)+4 \left(\sqrt{1-\sigma ^2}-1\right)+3 \mu (\sigma +1)
    \arcsin(\sigma)}} \nonumber \\
  && \qquad + \frac{2 \sqrt{3} \ln\left(\frac{\displaystyle 1-\mu }
    {\displaystyle -\mu +\sigma +\xi}\right)}{\sqrt{\frac{\displaystyle 2
    (-3 \mu +\sigma +2) \xi}{\displaystyle \mu  (2 \mu -\sigma -1)
    \ln\left(\displaystyle \frac{1-\mu }{\displaystyle -\mu +\sigma +\xi }
    \right)}+3}}-\pi ,
\label{eqpms0}
\end{eqnarray}
which still depends on the arbitrary parameter $\sigma$.

\begin{figure}
\begin{center}
\includegraphics[width=15cm]{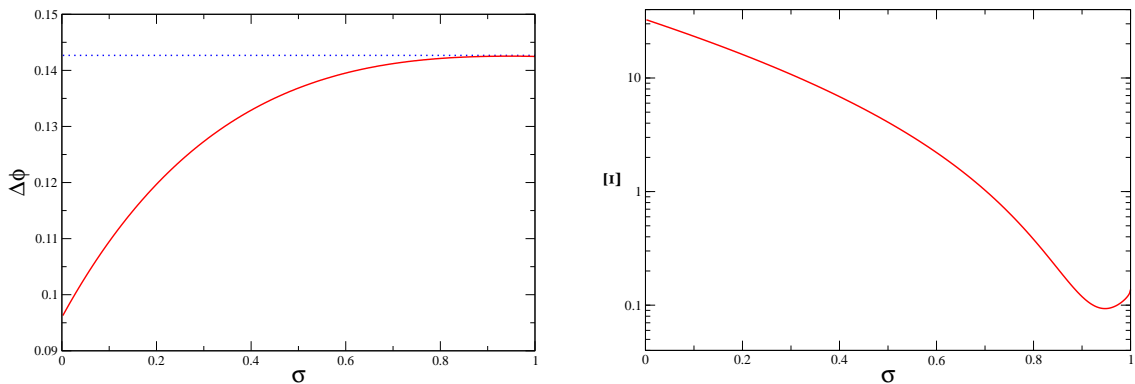}
\caption{ Left panel: approximate deflection angle for the Schwarzschild
  metric (Eq.~(\protect\ref{eqpms})) with $\mu = 10$ as a function
  of $\sigma$. The horizontal dotted line represents the exact value.
  Right panel: percent error of the approach as a function
  of $\sigma$.  (color online)}
\label{FIG2}
\end{center}
\end{figure}

\begin{figure}
\begin{center}
\includegraphics[width=15cm]{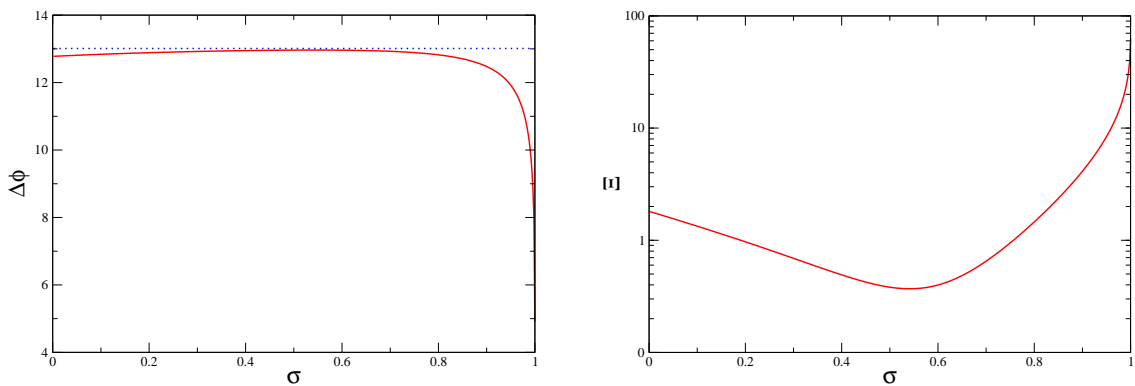}
\caption{ Same as in Fig.~\protect\ref{FIG2} for $\mu=1.001$.  (color online)}
\label{FIG3}
\end{center}
\end{figure}

Figs.~\ref{FIG2} and \ref{FIG3} show the
approximate deflection angle given by Eq.~(\ref{eqpms0}) and its percent error
$\Xi=100 \times\left|(\Delta\phi^{(1)}_{\text{PMS}}-\Delta\phi_{\text{exact}})/
\Delta\phi_{\text{exact}}\right|$ for two values of $\mu$.
We appreciate that the arbitrary parameter $\sigma$ can
also be determined by the PMS. In Fig.~\ref{FIG2}, $\mu = 10$ and
the optimal value of $\sigma$ is close to $1$, as expected. On the other hand,
if we take $\mu=1.001$, i.e. close to the photon sphere, the optimal value of
$\sigma$ drops to about $1/2$.
It is quite remarkable that in both cases the error 
made by choosing the optimal value for $\sigma$ 
is smaller than $1 \%$. {\sl Another important
observation is that the maximum of $\Delta\phi$ is quite flat and, consequently, 
a slightly imprecise estimation of $\sigma_{\text{PMS}}$ will not affect the 
precision of the approximation drastically}.

For this reason we do not pretend to obtain $\sigma$ directly by solving the
PMS condition, $\partial\Delta\phi^{(1)}/\partial\sigma=0$ (which is equivalent
to finding the maximum of the curve in the left panels of Fig.~\ref{FIG2} and 
\ref{FIG3}), since that would certainly be a difficult task and lead to quite 
involved expressions, but we rather use a simple analytical approximation,
which correctly describes the limits $\mu \rightarrow \infty$ and 
$\mu \rightarrow 1^+$. As noticed above, since the maximum is quite flat for 
$\mu \rightarrow 1^+$ one expects only a modest loss in precision, while
providing much simpler expressions.

\begin{figure}
\begin{center}
\includegraphics[width=9cm]{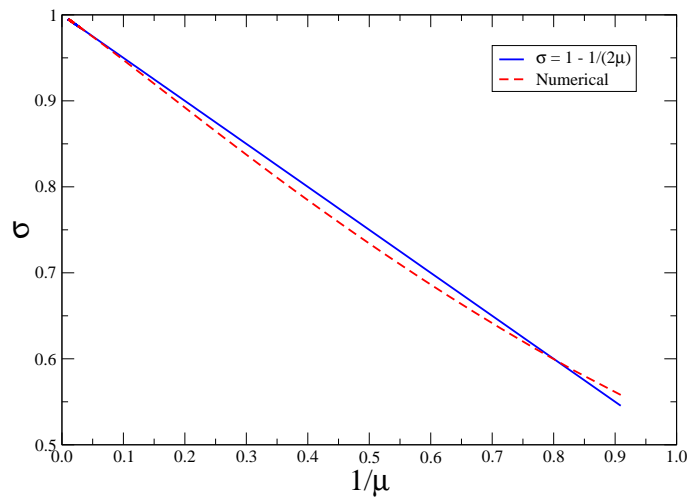}
\caption{ $\sigma_{\text{PMS}}$ obtained numerically solving the PMS condition (broken line) 
as a function of $1/\mu$  and the linear approximation $\sigma = 1 - 1/2\mu$ (solid line).  (color online)}
\label{FIG4}
\end{center}
\end{figure}

\begin{figure}
\begin{center}
\includegraphics[width=15cm]{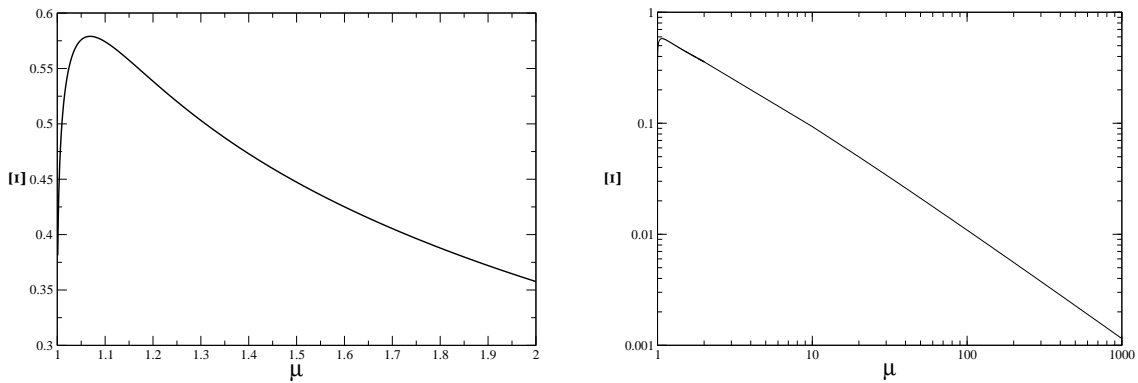}
\caption{Percent error of Eq.~(\protect\ref{eqpms}) as a function of $\mu$
for the deflection angle in the case of the Schwarzschild metric. Left panel: range $1\leq \mu\leq 2$; Right panel: range $1\leq \mu \leq 1000$.}
\label{FIG5}
\end{center}
\end{figure}

In Fig.~\ref{FIG4} we have plotted the exact value of $\sigma$ obtained solving 
numerically the PMS condition and the reasonable analytical approximation 
$\sigma_{\text{PMS}} \approx 1 - 1/2\mu$. 
Thus we obtain
\begin{equation}
  \Delta\phi_{\text{PMS}}^{(1)} = \frac{2\sqrt{6} \mu \left[\arcsin\left(1-
    \frac{\displaystyle 1}{\displaystyle 2 \mu}\right)\right]^{3/2}}
    {\sqrt{6 \arcsin\left(1-\frac{\displaystyle 1}{\displaystyle 2 \mu }\right)
    \mu ^2- 8\mu +\frac{\displaystyle 2 (6 \mu -1)}{\displaystyle
    \sqrt{4 \mu -1}}}}
   + \frac{2 \sqrt{6} \mu  \ln^{\frac{3}{2}}\left(
    \frac{\displaystyle \mu }{\displaystyle \mu -1}\right)}{\sqrt{6 \ln\left(
    \frac{\displaystyle \mu}{\displaystyle \mu -1}\right) \mu ^2-6 \mu +
    \frac{\displaystyle 1}{\displaystyle 2 \mu -1}+3}}-\pi ,
\label{eqpms}
\end{equation}
which provides an accuracy better than $1\%$ for {\em all} values of $\mu$, even
arbitrarily close to the photon sphere, as shown in Fig.~\ref{FIG5}.
Later on we will derive an even simpler analytical formula for the deflection angle
from Eq.~(\ref{eqpms}).

\section{General metric}
\label{gm}

We will now attack the problem of obtaining a first order formula for a general
static and spherically symmetric metric, in analogy with what has been done in
Refs.~\cite{AA06,AAF06}. For a given metric, once the functions $A(r)$, $B(r)$
and $D(r)$ are given (see Eq.~(\ref{eq_1})), one obtains a potential $V(z)$, as previously explained.
We assume that the potential admits two different expansions,
one at $z=0$, and one around $z=\mu$, which is a local maximum:
\begin{equation}
  V(z) = \sum_{n=1}^\infty v_n z^n = \sum_{n=0}^\infty \tilde{v}_n (z-\mu)^n .
\end{equation}

Notice that the first series runs from $n=1$, since $z=0$ is not necessarily a
local minimum of $V(z)$, although  this was the case for the Schwarzschild
metric. Clearly, in particular cases, such as the one previously examined, the
series coefficients vanish after a certain value of $n$, thus yielding
polynomial potentials.

Following the discussion of the preceding section we split the integral as
in Eq.~(\ref{eq:split}) and proceed to the calculation of each part.

\subsubsection{Region I ($0 \leq z \leq \sigma$)}

In this region we expand  $V(z)$ around $z=0$ and  use the interpolating
potential $V_a(z) = \lambda z^2$.
After expanding to first order we obtain Eq.~(\ref{eq:Da1}) with
\begin{equation}
  \Delta_a(z) =  \sum_{n=1}^\infty \frac{v_n}{\lambda} \sum_{k=0}^{n-1}
    \frac{z^k}{1+z} -1 .
\end{equation}
The deflection angle can now be written as
\begin{equation}
  \Delta\phi_a^{(1)} = \frac{3}{\sqrt{\lambda}}\arcsin\left(\sigma\right) -
    \frac{1}{\lambda^{3/2}} \sum_{n=1}^\infty v_n \ \sum_{k=0}^{n-1}
    I_k(\sigma),
\end{equation}
where we have defined the integrals
\begin{equation}
  I_{k}(\sigma) \equiv \int_0^{\arcsin\sigma} \frac{\sin^k\theta}{1+\sin\theta}
    d\theta
\end{equation}
which can be calculated exactly; for example:
\begin{subequations}
\label{eq:IntsI}
\begin{eqnarray}
  I_{0}(\sigma) &=& 1-\frac{\sqrt{1-\sigma }}{\sqrt{1+\sigma }} \\
  I_{1}(\sigma) &=& \arcsin(\sigma )+\frac{\sqrt{1-\sigma }}{\sqrt{1+\sigma}}-1
    \\
  I_{2}(\sigma) &=& -\arcsin(\sigma )-\sqrt{1-\sigma ^2}-\frac{\sqrt{1-\sigma}}
    {\sqrt{1+\sigma }}+2 \\
  I_{3}(\sigma) &=& \frac{3}{2} \arcsin(\sigma)  - 2 + \frac{1}{2}
    \left(-\sigma ^2+\sigma +4\right) \sqrt{\frac{2}{\sigma +1}-1} \\
  I_{4}(\sigma) &=& -\frac{3}{2} \arcsin(\sigma) + \frac{8}{3} + \frac{1}{6}
    \sqrt{\frac{1-\sigma }{\sigma +1}} \left(-2 \sigma ^3+\sigma^2-7 \sigma
    -16\right) .
\end{eqnarray}
\end{subequations}
The PMS yields
\begin{equation}
  \lambda_{\text{PMS}} =  \dfrac{\  \sum_{n=1}^\infty v_n \ \sum_{k=0}^{n-1}
    I_k(\sigma)}{\arcsin\left(\sigma \right)} ,
\end{equation}
so that
\begin{equation}
  \Delta\phi_a^{(1)} = \dfrac{{2 \arcsin^{3/2}\sigma}}{\sqrt{ \sum_{n=1}^\infty
    v_n \sum_{k=0}^{n-1} I_k(\sigma)}}.
\end{equation}
Taking into account the form of the integrals in Eq.~(\ref{eq:IntsI}) we can
express the deflection angle as
\begin{equation}
\label{eq_33}
  \Delta\phi_a^{(1)} = \dfrac{\arcsin^{3/2}\left(\sigma \right)}{\sqrt{
    F_1(\sigma) + F_2 \arcsin\sigma}}.
\end{equation}

\subsubsection{Region II ($\sigma \leq z \leq 1$)}

We now come to the second region where the potential is expressed in terms of a
series around  the local maximum at $\mu$.
After expanding to first order, and using the same interpolating potential as in
the case of the Schwarzchild metric $V_b(z) = V(\mu)-\rho (z-\mu)^2$, we
obtain
\begin{equation}
  \Delta\phi_b^{(1)} = \frac{2}{\sqrt{\rho}} \int_{\sigma}^1
    \frac{dz}{\sqrt{(z-\mu)^2-(1-\mu)^2}} \left[ 1 - \frac{ \Delta_b(z)}{2}
    \right] ,
\end{equation}
where
\begin{equation}
  \Delta_b(z) \equiv \frac{V(1)-V(z)}{V_b(1)-V_b(z)}-1  = \sum_{n=2}^\infty
    \frac{\tilde{v}_n (a^n-b^n)}{-\rho (a^2-b^2)} - 1 ,
\end{equation}
and
\begin{equation}
  a \equiv 1-\mu , \quad b \equiv z-\mu .
\end{equation}
Since
\begin{equation}
  a^n -b^n = (a-b) \sum_{k=0}^{n-1} a^k b^{n-1-k}
\end{equation}
we have
\begin{equation}
  \Delta_b(z) \equiv \frac{V(1)-V(z)}{V_b(1)-V_b(z)}-1  = \sum_{n=2}^\infty
    \frac{\tilde{v}_n \sum_{k=0}^{n-1} a^k b^{n-1-k} }{-\rho (a+b)} - 1 .
\end{equation}
In terms of the new variable $u = b/a$ one obtains
\begin{equation}
  \Delta\phi_b^{(1)} = \frac{2}{\sqrt{\rho}} \int_{u_-}^{u_+}
  \frac{du}{\sqrt{u^2-1}}  \left[ 1 - \frac{ \Delta_b(z(u))}{2} \right] ,
\end{equation}
where $u_+ \equiv (\mu-\sigma)/(\mu-1)$ and $u_- \equiv 1$.

Notice that
\begin{equation}
  \frac{a^k b^{n-1-k} }{(a+b)} = \frac{u^{n-1-k}}{1+u} \ (1-\mu)^{n-2} .
\end{equation}
Therefore we get
\begin{equation}
  \Delta\phi_b^{(1)} = \frac{2}{\sqrt{\rho}} \int_{u_-}^{u_+}
    \frac{du}{\sqrt{u^2-1}} \left[ \frac{3}{2} - \frac{1}{2}\sum_{n=2}^\infty
    \frac{\tilde{v}_n}{-\rho(1+u)} \sum_{k=0}^{n-1} u^{n-1-k} (1-\mu)^{n-2}
    \right] ,
\end{equation}
where we have defined
\begin{equation}
  J_{k}(\sigma)=\int_{u_{-}}^{u_{+}}\dfrac{u^{k}}{\sqrt{u^{2}-1}(u+1)}du .
\end{equation}
The first integrals are
\begin{subequations}
\label{eq:IntsII}
\begin{eqnarray}
  J_{0}(\sigma) &=& -\frac{\epsilon-1}{\epsilon +1} \\
  J_{1}(\sigma)&=& \frac{\epsilon -1}{\epsilon +1}-\ln (\epsilon ) \\
  J_{2}(\sigma) &=&  -\dfrac{\epsilon-1}{(\epsilon+1)2\epsilon}
    (1+4\epsilon+\epsilon^{2})+\ln(\epsilon) \\
  J_{3}(\sigma)&=& \frac{\epsilon-1 }{8 \epsilon^2 (1+\epsilon)} \left(
    -\epsilon^4+2 \epsilon^3+14 \epsilon^2+2 \epsilon -1\right)
    -\frac{3}{2} \ln ( \epsilon ) \\
  J_{4}(\sigma)&=& \dfrac{\epsilon-1}{24\epsilon^{3}(\epsilon+1)}
    \left( -\epsilon^6+\epsilon^5-17 \epsilon^4-62 \epsilon^3-17
    \epsilon^2+\epsilon -1 \right) +  \frac{3}{2}  \ln (\epsilon ) ,
\end{eqnarray}
\end{subequations}
where
\begin{equation}
  \epsilon \equiv \frac{\mu -\sigma +\sqrt{\sigma ^2-2 \mu  \sigma +2 \mu
    -1}}{\mu -1} .
\end{equation}
The PMS yields
\begin{equation}
  \rho_{\text{PMS}}=\frac{1}{\ln\epsilon} \sum_{n=2}^\infty
    \tilde{v}_n(1-\mu)^{n-2}\sum_{k=0}^{n-1}J_{k}(\sigma)
\end{equation}
and we obtain
\begin{equation}
  \Delta\phi_b^{(1)} =\frac{2 \ln^{3/2}\epsilon}{\sqrt{\sum_{n=2}^\infty
    \tilde{v}_n(1-\mu)^{n-2}\sum_{k=0}^{n-1}J_{k}(\sigma)}}.
\end{equation}
Once again, looking at the structure of the integrals (\ref{eq:IntsII}) we write
\begin{equation}
\label{eq_47}
  \Delta\phi_b^{(1)} =\frac{\ln^{3/2}\epsilon}{\sqrt{G_1(\sigma)+G_2\
    \ln\epsilon}} \ .
\end{equation} 

Notice that the explicit expression of the coefficients $F_{1,2}$ and $G_{1,2}$
will depend on the metric.

\section{results}
\label{sec:results}

\subsection{Schwarzschild metric}

Our first application is to the Schwarzschild metric, which corresponds to
\beq
B(r) = A^{-1}(r) = \left(1-\frac{2GM}{r}\right)  \ \ , \ \ D(r) = 1 \ .
\eeq
Here $M$ is the Schwarzschild mass. The angle of deflection of a ray of light reaching a minimal distance
$r_0$ from the black hole can be obtained using Eq.~(\ref{eq_1_5}).
The exact result can be expressed in terms of incomplete elliptic integrals of
the first kind \cite{Darw59}
and reads 
\beq
\Delta\phi = 4 \sqrt{\frac{\overline{r}_0}{\Upsilon}} \ \left[ F\left(\frac{\pi}{2}, \kappa\right) - 
F\left(\varphi, \kappa\right)  \right] \ ,
\label{darwin}
\eeq
where $\overline{r}_0\equiv r_0/GM$ and
\beq
\Upsilon \equiv \sqrt{\frac{\overline{r}_0-2}{\overline{r}_0+6}} \ \ &,& \ \  
\kappa \equiv \sqrt{(\Upsilon-\overline{r}_0+6)/2\Upsilon} \ \ , \ \ 
\varphi \equiv \sqrt{\arcsin \left[\frac{2+\Upsilon-\overline{r}_0}{6+\Upsilon-\overline{r}_0}\right]} \ .
\eeq

We compare our analytical formulas with the exact one and with the approximation proposed by Bozza \cite{Boz02}:
\begin{equation}
  \Delta\phi_{\text{Bozza}} = -2 \ln \left[\frac{1}{12} \left(2+\sqrt{3}\right)
    (\mu-1)\right]-\pi.
\end{equation}
For brevity, we shall refer to the approximation developed in the previous Section as Method~I. In the 
Appendix we have also derived a simpler analytical expression for the deflection angle using an alternative
method, which we refer to as being Method~II.

\begin{figure}
\begin{center}
\includegraphics[width=16cm]{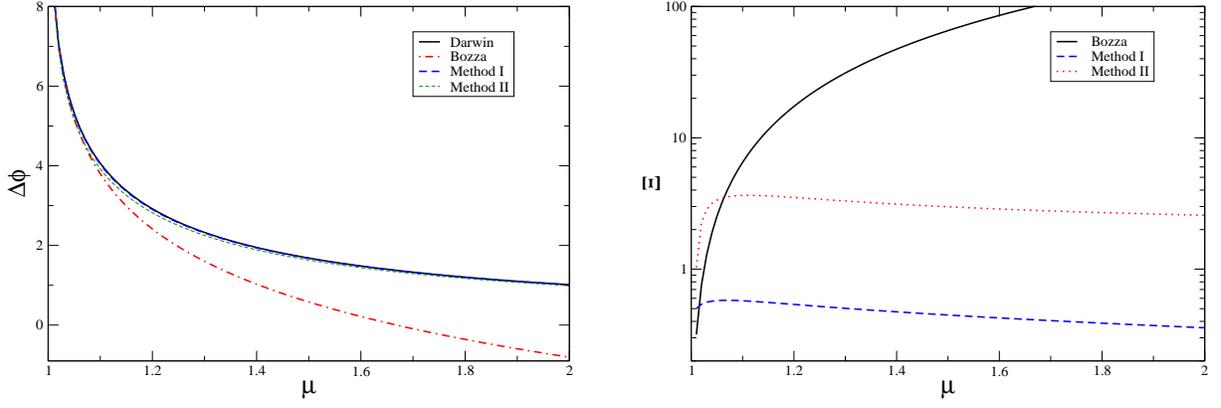}
\caption{Left panel: deflection angle for the Schwarzschild metric as a
  function of $\mu$ using different approximations and the exact
  result. Right panel: percent error of the approximate solutions as a
  function of $\mu$.  (color online)}
\label{FIG6}
\end{center}
\bigskip
\end{figure}

\begin{figure}
\begin{center}
\includegraphics[width=8cm]{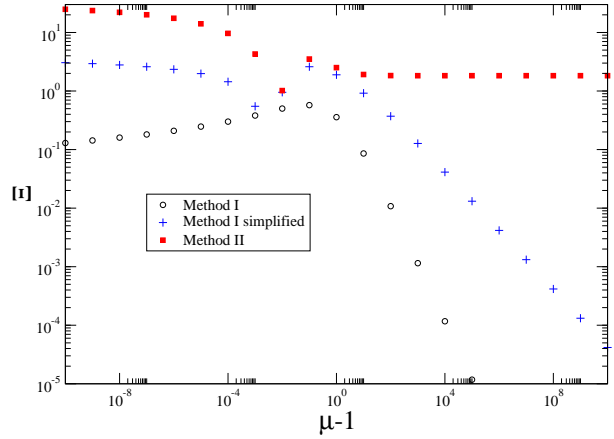}
\caption{ Percent error for the deflection angle in the Schwarzschild
  metric calculated with Method I (Eq.~(\protect\ref{eqpms})), with the
  simplified version of Method I (Eq.~(\protect\ref{eq:DSsimp}) and with
  Method II (Eq.~(\protect\ref{eq:Dphi_ap_S})).  (color online)}
\label{FIG6a}
\end{center}
\bigskip
\end{figure}

In the former approach, the expression of Eq.~(\ref{eqpms}) can be
further simplified mantaining good accuracy by observing that
\begin{subequations}
\begin{equation}
  \frac{2\sqrt{6}\mu\arcsin^{\frac{3}{2}}\left(1-\frac{\displaystyle
    1}{\displaystyle 2 \mu}\right)}{\sqrt{6\arcsin\left(1-
    \frac{\displaystyle 1}{\displaystyle 2 \mu }\right) \mu^2 - 8\mu +
    \frac{\displaystyle 2 (6 \mu - 1)}{\displaystyle \sqrt{4 \mu -1}}}}
    \approx \pi +\frac{12}{-3 \sqrt{4 \mu -1}-4}
\end{equation}
and
\begin{equation}
  \frac{2 \sqrt{6} \mu \ln^{\frac{3}{2}}\left(\frac{\displaystyle \mu }
    {\displaystyle \mu -1}\right)}{\sqrt{6 \ln \left(\frac{\displaystyle \mu}
    {\displaystyle \mu -1}\right) \mu^2 - 6 \mu +\frac{\displaystyle 1}
    {\displaystyle 2 \mu -1}+3}} \approx \sqrt{4\mu+1/3} \ln
    \left(\frac{\mu}{\mu -1}\right) .
\end{equation}
\end{subequations}
Therefore we get the much simpler expression
\begin{equation}
\label{eq:DSsimp}
  \Delta\phi_{\text{PMS}}^{(1)} \approx \frac{12}{-3 \sqrt{4 \mu -1}-4} +
    \sqrt{4\mu+1/3} \ln \left(\frac{\mu}{\mu -1}\right).
\end{equation}
In Fig.~\ref{FIG6} we compare our approximations, the one of Bozza \cite{Boz02},
 and the exact result of Eq.(\ref{darwin}).
It is clear that --- apart from the region very close to $\mu=1$ --- our
approximations are more accurate, especially as $\mu$ increases: the curves
in the left plot corresponding to using our approximations are hardly 
distinguishable from the exact one. 
Indeed, in the case of Method I, after having reached a maximum error close to $\mu=1$, 
the precision keeps improving, as one can see in Fig.~\ref{FIG6a}. In the case of Method II
and of the simplified expression of Eq.~(\ref{eq:DSsimp}) the error does not tend to 
zero as the photon  sphere is approached but it appears to remain finite.

\subsection{Reissner-Nordstr\"om metric}

The Reissner-Nordstr\"om (RN) metric describes a black hole with charge and corresponds to 
\beq
B(r) = A^{-1}(r) =  \left(1-\frac{2 GM}{r}+\frac{q^2}{r^2}\right) \  \  &,& \ \ D(r) = 1 \ .
\eeq
The corresponding potential is found to be
\begin{eqnarray}
  V(z) &=& z^2 \left(\frac{q^2}{r_0^2}z^2-2\frac{M}{r_0}z+1\right)
  = z^2  - \frac{2M}{\mu r_\text{PS}} z^3 + \frac{q^2}{\mu^2  r_\text{PS}^2} z^4,
\end{eqnarray}
where $r_\text{PS} = 4 q^2/(3 M - \sqrt{9 M^2 - 8 q^2})$ is the photon sphere
and $\mu = r_0/r_\text{PS}$; $q$ is the charge of the black hole.
Also in this case we have found that $\sigma = 1-1/2\mu$ is a satisfactory
analytical approximation to the optimal value $\sigma_{\text{PMS}}$. Although the 
exact solution of the PMS condition would in general depend on the charge $q$, 
we have verified that the general features discussed in the case of the Schwarzchild metric
remain valid and only a quite limited error is introduced by this choice.

\begin{table}
\caption{\label{tab1} Numerical values of the coefficients $A$ and $B$:
$A_\text{Eiroa}$ and $B_\text{Eiroa}$ are taken from Table~1 of
  Ref.~\protect\cite{Eir02}, whereas $A_\text{us}$ and $B_\text{us}$ are
  obtained from Eq.~(\protect\ref{eq:AB}).}
\begin{ruledtabular}
\begin{tabular}{ccccccc}
$|q|$ & $0$ & $0.1 \ M$  & $0.25 \ M$ & $0.5 \ M$  & $0.75 \ M$ & $1 \ M$ \\
\hline
$A_\text{Eiroa}$ & 2.00000  & 2.00224  & 2.01444 & 2.06586  & 2.19737  &
  2.82843 \\
$A_\text{us}$    & 2.00000  & 2.00224  & 2.01444 & 2.06586  & 2.19737  &
  2.82843 \\
$B_\text{Eiroa}$ & 0.207338 & 0.207979 & 0.21147 & 0.225997 & 0.262085 &
  0.426782 \\
$B_\text{us}$    &  0.213892 & 0.214535  & 0.218032  & 0.232554  & 0.268419  &
  0.430856 \\
\end{tabular}
\end{ruledtabular}
\end{table}

Using the former we obtain
\begin{subequations}
\begin{eqnarray}
  F_1 &=& \frac{-4 \alpha^3+6 \alpha^2-4 \alpha +2}{3 \alpha
    \left(\alpha^2+1\right)^2} + \frac{-2 \alpha^7+3 \alpha^6-6 \alpha^5+4
    \alpha^4-6 \alpha^3+8 \alpha^2-2 \alpha +1}{12 \alpha \left(\alpha^2+1
    \right)^2} \frac{q^2}{r_0^2}  \\
  F_2 &=& \frac{1}{4}+\frac{3 q^2}{8 r_0^2} ,
\end{eqnarray}
\end{subequations}
where $\alpha \equiv \sqrt{4 \mu-1}$, and
\begin{subequations}
\begin{eqnarray}
  G_1 &=& \frac{-3 \alpha^4+6 \alpha^2+1}{3 \left(\alpha^2-1\right)
    \left(\alpha^2+1\right)^2} + \frac{\left(21 \alpha^8+12 \alpha^6-110
    \alpha^4-28 \alpha^2-23\right)}{96 \left(\alpha^2-1\right)
    \left(\alpha^2+1\right)^2} \frac{q^2}{r_0^2} \\
  G_2 &=&  \frac{1}{4} -  \frac{\left(7 \alpha^4-10 \alpha^2+31\right) }{128}
    \frac{q^2}{r_0^2}, 
\end{eqnarray}
\end{subequations}
where the coefficients $F_{1,2}$ and $G_{1,2}$ were previously introduced in 
the previous section (see Eqs.~(\ref{eq_33}) and (\ref{eq_47})).

Using our expression for the deflection angle we have obtained the coefficients
$A$ and $B$ introduced by Eiroa et al. in Ref.~\cite{Eir02}:
\beq
\Delta\phi \approx - A \log \left( B  \ \epsilon \right) - \pi,
\eeq
where $\epsilon = r_0 - r_{\text{ps}}$, $r_{\text{ps}}$ being the distance
corresponding to the photon sphere.

We have found
\begin{subequations}
\label{eq:AB}
\begin{eqnarray}
  A &=& - \lim_{\mu\rightarrow 1^+}
    \frac{\Delta\phi_\text{PMS}^{(1)}}{\log\left(\mu-1\right)} =
    \frac{4 q/M}{\sqrt{8 q^2/M^2+3 \sqrt{9-8 q^2/M^2}-9}} \\
  B &=&  \frac{4}{\psi+3} \exp[- 0.234 + 0.203/\psi - 1.096
    \sqrt{\psi/(\psi+0.826)}] ,
\end{eqnarray}
\end{subequations}
where $\psi \equiv \sqrt{9-8 q^2/M^2}$.

Table~\ref{tab1} shows that our analytical formulas are in remarkable agreement
with the numerical results of Eiroa et al.~\cite{Eir02}. In particular our
expression for the coefficient $A$ appears to be exact.

\begin{figure}
\begin{center}
\includegraphics[width=9cm]{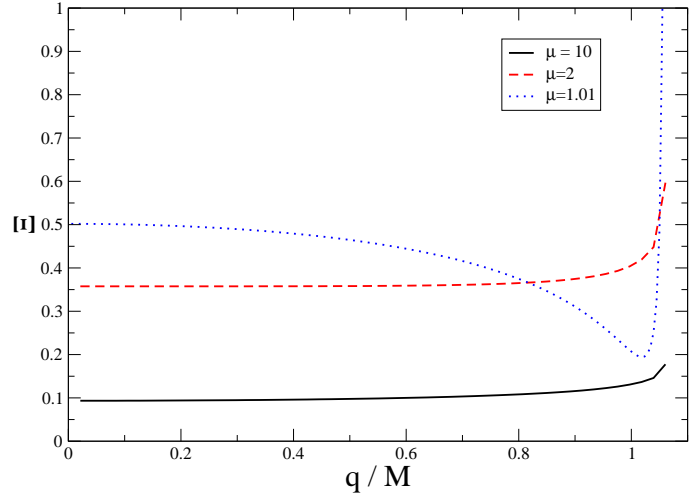}
\caption{ Percent error of the deflection angle in the
  Reissner-Nordstr\"om metric calculated with Method I at different values of
  $\mu$.  (color online)}
\label{FIG7}
\end{center}
\end{figure}

Fig.~\ref{FIG7} displays the percent error of our approach for different values
of $\mu$ as a function of $q/M$. Again, the error is generally below 0.5\% and
gets smaller as $\mu$ increases.

Notice that we do not need to discuss the WDL of our formulas, since in this case
one has that $\mu \rightarrow \infty$ and the method of \cite{AA06,AAF06} is recovered.
The reader will find  a detailed comparison of our method with other methods available 
in the literature in \cite{AA06,AAF06}.

\subsection{Janis-Newman-Winicour metric}

Finally, we consider the spherically symmetric metric solution to the Einstein
massless scalar equations~\cite{JNW68},
\begin{equation}
  A(r) = \left(1-b/r\right)^{-\nu}, \quad B(r) = \left(1-b/r\right)^{\nu},
    \quad D(r) = \left(1-b/r\right)^{1-\nu},
\end{equation}
which reduces to the Schwarzschild metric for $\nu =1$ and $b = GM$.
For this metric we obtain the potential
\begin{equation}
  V(z) = -\left(1-\frac{b}{r_0}\right)^{2 \nu -1} \left(1-\frac{b z}
    {r_0}\right)^{2-2 \nu }+z^2 \left(1-\frac{b z}{r_0}\right)
    +\left(1-\frac{b}{r_0}\right)^{2 \nu -1},
\end{equation}
that can be expanded around $z=0$ to give
\begin{eqnarray}
  V(z) &\approx& v_1 z + v_2 z^2 + v_3 z^3 + \dots \nonumber \\
  &\approx& -2 (\nu -1) (1-\frac{b}{r_0})^{2 \nu -1} \frac{b}{r_0} z +
    \left[1-(\nu -1) (2 \nu -1) \left(1-\frac{b}{r_0}\right)^{2 \nu -1}
    \left(\frac{b}{r_0}\right)^2\right] \ z^2  \nonumber \\
  &&\quad + \left[-\frac{2}{3} (\nu -1) \nu  (2 \nu -1)
    \left(\frac{b}{r_0}\right)^3 \left(1-\frac{b}{r_0}\right)^{2 \nu-1}
    -\frac{b}{r_0}\right] \ z^3 + O\left[z^4\right] .
\end{eqnarray}
Notice that the radius of convergence of the series of $V(z)$ 
around $z=0$ is $\bar{z} = r_0/b$. We therefore ask that $\bar{z} \geq 1$, 
i.e. that $r_0\geq b$. Clearly the accuracy of the expansion above
will depend upon the location of $\bar{z}$ and a larger number of
terms is expected to be needed when $\bar{z}$ approaches one.

\begin{figure}
\begin{center}
\includegraphics[width=15cm]{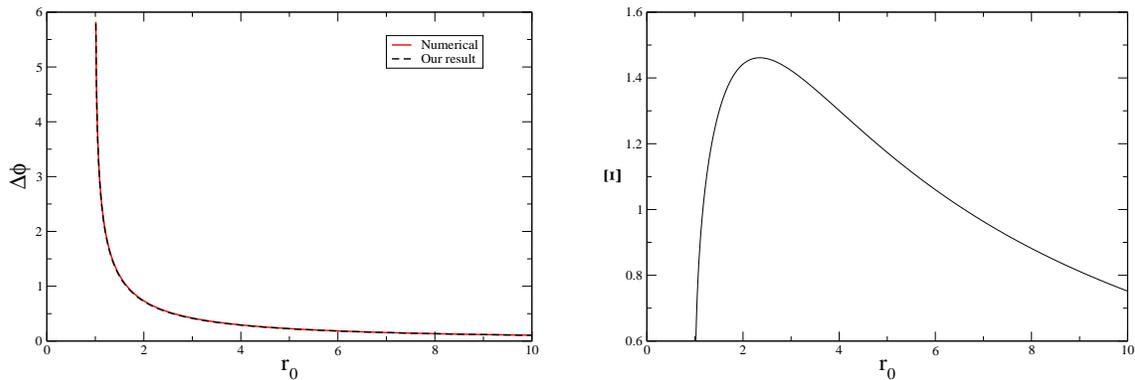}
\caption{ Left panel: deflection angle in the Janis-Newman-Winicour metric
  using $\nu = 1/2$ and $b=1$; the solid line is the numerical result, whereas
  the dashed line corresponds to our analytical formula. Right panel:
  percent error for the deflection angle using our analytical formula. }
\label{FIGJNW}
\end{center}
\end{figure}

We then find
\begin{subequations}
\begin{eqnarray}
  F_1 &=&  \frac{(\alpha-1) (1+\alpha^2)}{8 \alpha} v_2 +
    \frac{(\alpha -1) \left(\alpha  \left(3 \alpha ^5+9 \alpha ^3-23 \alpha
    +16\right)-13\right)}{64 \left(\alpha ^3+\alpha \right)} v_3 \\
  F_2 &=& \frac{v_2}{4} \\
  G_1 &=&  \frac{(6 (\mu -1) \mu +1)}{8 \mu  (2 \mu -1)}  v_3 \\
  G_2 &=&  -\frac{v_2+3\mu v_3}{4} ,
\end{eqnarray}
\end{subequations}
where, again, $\alpha \equiv \sqrt{4 \mu-1}$ and $F_{1,2}$ and $G_{1,2}$ have
been introduced in Eqs.~(\ref{eq_33}) and (\ref{eq_47}). 
Notice that in the cubic approximation we can express the coefficient $v_1$ in
terms of the other parameters:
\begin{equation}
  v_1 = -2\mu v_2 -3\mu^2 v_3.
\end{equation}
Fig.~\ref{FIGJNW} compares our approximation for the deflection angle with
the exact result. One can see that also for this metric the accuracy is very
good.

\section{Conclusions}

We have presented a new method for obtaining analytical expressions for
the deflection angle of light in a static and spherically symmetric metric,
which is accurate in both the weak and strong regimes. The former corresponding
to lensing at arbitrarily large distances from the compact body and the latter
to distances arbitrarily close to the photon sphere. Our first--order
analytical formulas exhibit errors below $1 \%$ at {\sl all} distances
from the compact body. For this reason, although our method can be
applied to any given order, in a way similar to what has been done in
Ref.~\cite{AAF06} with the method of Ref.~\cite{AA06}, the accuracy of our
first--order expressions is certainly sufficient for most physical applications.

Moreover, the method that we have presented in this paper reduces to the
previous method discussed in Refs.~\cite{AA06,AAF06} in the weak deflection
limit (WDL), since the arbitrary parameter $\sigma$ tends to unity in this
limit. For this reason, for the  
comparison with alternative methods developed to describe the WDL we have relied
on the discussion contained in \cite{AA06,AAF06}. 

To the best of our knowledge, our method is the only one available that 
allows one to obtain completely analytical formulas which are valid in both
SDL and WDL regimes, regardless of the particular static and spherically symmetric 
metric used. The results that we obtain are clearly non--perturbative, since they 
do not correspond to a polynomial expression in any of the physical parameters in
the model and provide the correct logarithmic strength of the singularities. 
Moreover, our analytical expressions never involve special functions 
and are easy to evaluate. Just to mention one success of our approach, in the case of
the Reissner-Nordstrom metric, we have obtained an analytical formula for the coefficients
$A$ and $B$ which have been numerically calculated by Eiroa et al. in Ref.~\cite{Eir02}: in the
case of the coefficient $A$ our formula reproduces {\sl all the digits} given
by the numerical calculation.
We also wish to mention that our method relies on solid mathematical grounds and could be
used to obtain an exact series representation for the integrals of the deflection angle:
we have not considered this issue because of the highly precise results that are obtained
already working to order one.

Finally, we have also discussed --- showing an application to the Schwarzschild metric
--- an alternative method that yields larger errors but even simpler expressions and 
that can be useful in cases where a somewhat reduced accuracy can be traded for the 
possibility of more convenient analytical manipulations.

\appendix 

\section{Alternative formula through linear interpolation}
\label{subsec:linint}

In this appendix we present an alternative method to obtain the deflection angle in 
a given static and spherically symmetric metric. This method is a generalization of a 
recently published approach to the period of the simple pendulum~\cite{Lima06}. 
It is our purpose to obtain a simple analytical approximation to an integral of the form
\begin{equation}
I=\int_{0}^{1}\frac{dz}{\sqrt{Q(z)}}  \label{eq:int_I_z}
\end{equation}
where $Q(1)=0$. We assume that any other zero of $Q(z)$ is outside the
closed interval $[0,1]$.

We define a reference function
\begin{equation}
Q_{0}(z)=(1-z)(z-z_{0}),  \label{eq:Q0}
\end{equation}
where $z_{0}<0$, and carry out the change of variables
\begin{equation}
z=\frac{1+z_{0}}{2}+\frac{1-z_{0}}{2}\sin \theta .  \label{eq:z(theta)}
\end{equation}
We thus obtain
\begin{eqnarray}
I &=&\int_{\theta _{0}}^{\pi /2}\frac{d\theta }{\sqrt{F(z(\theta ))}}
,\quad\theta _{0}=\arcsin \left( \frac{1+z_{0}}{z_{0}-1}\right), \nonumber \\
F(z) &=&\frac{Q(z)}{Q_{0}(z)}.  \label{eq:int_I_theta}
\end{eqnarray}
We next substitute a linear funcion $\alpha +\beta \theta $ for $\sqrt{%
F(z(\theta ))}$ such that $\sqrt{F(0)}=\alpha +\beta \theta _{0}$, $\sqrt{%
F(1)}=\alpha +\beta \pi /2$ and obtain the general approximate expression
\begin{equation}
I\approx\frac{\pi -2\theta _{0}}{4\left( \sqrt{F(1)}-\sqrt{F(0)}\right) }\ln
\left( \frac{F(1)}{F(0)}\right) .  \label{eq:I_ap_gen}
\end{equation}
We have yet to specify the exact location of $z_{0}$. Notice that
\begin{equation}
F(1)=\lim_{z\rightarrow 1}\frac{Q(z)}{(1-z)(z-z_{0})}=\frac{Q^{\prime }(1)}{%
z_{0}-1}.  \label{eq:F(1)}
\end{equation}
In the case of the Schwarzschild metric we have $\Delta \phi =\sqrt{6\mu }%
I-\pi $, where $I$ is given by Eq. (\ref{eq:int_I_z}), with
\begin{eqnarray}
Q(z) &=&(z-z_{1})(z-z_{2})(z-z_{3}),  \nonumber \\
z_{1} &\leq &0\leq z_{3},\;z_{2}=1,  \label{eq:Q(z)_S}
\end{eqnarray}
and $z_{3}\geq 1$ if $\mu \geq 1$.

We may set the location of $z_{0}$ to have either the most accurate
analytical expression or the simplest one; in what follows we choose the
latter. If $z_{0}=-1$, then $\theta _{0}=0$ and
\begin{equation}
\Delta \phi \approx \frac{\sqrt{3\mu }\pi }{2}\left( \sqrt{3\mu -2}+\sqrt{
3\mu -3}\right) \ln \frac{3\mu -2}{3\mu -3}-\pi .  \label{eq:Dphi_ap_S}
\end{equation}
Notice the logarithmic singularity at $\mu =1$ that comes from the fact that
$z_{3}(\mu =1)=z_{2}=1$ and the integral diverges as $\mu \rightarrow 1^{+}$.
This approach is considerably less accurate than the preceding one, but we
have decided to include it in this paper for two reasons: first, it provides
simple and general expressions; second, its error is quite uniform.


\begin{thebibliography}{99}

\bibitem{Bray:1985ew} I.~Bray, Phys.\ Rev.\  D {\bf 34}, 367 (1986).
\bibitem{Sereno:2003kx} M.~Sereno,  Mon.\ Not.\ Roy.\ Astron.\ Soc.\  {\bf 344}, 942 (2003)
\bibitem{Sereno:2006ss}  M.~Sereno and F.~De Luca, Phys.\ Rev.\  D {\bf 74}, 123009 (2006)
\bibitem{Sereno:2003nd} M.~Sereno,  Phys.\ Rev.\  D {\bf 69}, 023002 (2004)
\bibitem{Keet05} C. R. Keeton and A. O. Petters, Phys. Rev. D {\bf 72}, 104006 (2005).
\bibitem{FKN00} S. Frittelli, T..P. Kling, and T. Newman, 
                Phys. Rev. D {\bf 61}, 064021 (2000).
\bibitem{VE00}  K. S. Virbhadra and G. F. R. Ellis, 
                Phys. Rev. D {\bf 62}, 084003 (2000).
\bibitem{VE02}  K. S. Virbhadra and G. F. R. Ellis, 
                Phys. Rev. D {\bf 65}, 103004 (2002).
\bibitem{VNC98} K. S. Virbhadra, D. Narasimha, and S. M. Chitre, 
                Astron. Astrophys. {\bf 337}, 1 (1998).
\bibitem{Eir02} E. F. Eiroa, G. E. Romero and D. F. Torres, 
                Phys. Rev. D {\bf 66}, 024010 (2002).
\bibitem{Bha03} A. Bhadra, Phys. Rev. D {\bf 67}, 103009 (2003).
\bibitem{Boz03} V. Bozza, Phys. Rev. D {\bf 67}, 103006 (2003);
                V. Bozza, F. De Luca, G. Scarpetta, and M. Sereno, 
                Phys. Rev. D {\bf 72}, 08300 (2005);
                V. Bozza, F. De Luca, and G. Scarpetta, 
                Phys. Rev. D {\bf 74}, 063001 (2006).
\bibitem{Whi05} R. Whisker, Phys. Rev. D {\bf 71}, 064004 (2005).
\bibitem{Eir05} E. F. Eiroa, Phys. Rev. D {\bf 71}, 083010 (2005).
\bibitem{Eir06} E. F. Eiroa, Phys. Rev. D {\bf 73}, 043002 (2006).
\bibitem{SB06}  K. Sarkar and A. Bhadra, 
                Class. Quant. Grav. {\bf 23}, 6101 (2006).
\bibitem{Kono06} R.~A.~Konoplya, Phys.\ Lett.\  B {\bf 644}, 219 (2007)
\bibitem{Gyulchev:2006zg} G.~N.~Gyulchev and S.~S.~Yazadjiev, Phys.\ Rev.\  D {\bf 75}, 023006 (2007).
\bibitem{Perl04} V. Perlick, Phys. Rev. D {\bf 69}, 064017 (2004).
\bibitem{BS06} V.~Bozza and M.~Sereno, Phys.\ Rev.\  D {\bf 73}, 103004 (2006)
\bibitem{CVE} C.M. Claudel, K.S. Virbhadra, G.F.R. Ellis,J.Math.Phys.42:818-838,2001.
\bibitem{Mutka} P. T. Mutka and P. M\"ah\"onen, 
                The Astrophysical Journal {\bf 581}, 1328 (2002);
                P. T. Mutka and P. M\"ah\"onen, 
                The Astrophysical Journal {\bf 576}, 107 (2002).
\bibitem{Belo02} A. M. Beloborodov, 
                 The Astrophysical Journal {\bf 566}, L85 (2002).
\bibitem{Boz02} V. Bozza, Phys. Rev. D {\bf 66}, 103001 (2002).
\bibitem{IP06}  S.~V.~Iyer and A.~O.~Petters,  arXiv:gr-qc/0611086.

\bibitem{AA06}  P. Amore and S. Arceo,
                Phys. Rev. D {\bf 73}, 083004 (2006).
\bibitem{AAF06} P. Amore, S. Arceo, and F. Fern\'andez,
                Phys. Rev. D {\bf 74}, 083004 (2006).
%\bibitem{Bozza02} V. Bozza,
%                  Phys. Rev. D {\bf 66}, 103001 (2002).
\bibitem{Am05a} P. Amore and R. A. Sa\'enz,
                Europhys. Lett. {\bf 70}, 425 (2005).
\bibitem{Am05b} P. Amore, A. Aranda, F. Fern\'andez, and R. A. Sa\'enz,
                Phys. Rev. E {\bf 71}, 016704 (2005).
\bibitem{Ste81} P. M. Stevenson, Phys. Rev. D {\bf 23}, 2916 (1981).


\bibitem{Darw59} C. Darwin, Proc.R. Soc. London {\bf A} 249, 180 (1959); 
                 C. Darwin, Proc.R. Soc. London {\bf A} 263, 39 (1961)

\bibitem{JNW68} A. I. Janis, E. T. Newman, and J. Winicour,
                Phys. Rev. Lett. {\bf 20}, 878 (1968);
                K.S. Virbhadra, Int.J.Mod.Phys.A12:4831-4836,1997. 
\bibitem{Lima06} F.M.S. Lima and P. Arun, Am. J. Phys. {\bf 74}, 892 (2006).
\end{thebibliography}
\end{document}